\documentclass[11pt,a4paper]{article}
\usepackage[left=0.5in,right=0.5in,top=0.5in,bottom=0.5in]{geometry}
\usepackage{amsmath,amssymb,amsthm,float}
\usepackage{tikz}
\usepackage{tkz-euclide}
\usetikzlibrary{arrows}
\tikzset{>=stealth}
\newtheorem{theorem}{Theorem}

\newcommand{\cT}{{\mathcal T}}
\newcommand{\cF}{{\mathcal F}}
\newcommand{\cK}{{\mathcal K}}
\newcommand{\cI}{{\mathcal I}}
\newcommand{\cJ}{{\mathcal J}}

\newcommand{\II}{I\kern -0.7ex I}
\newcommand{\III}{I\kern -0.7ex I\kern -0.7ex I}

\newcommand{\be}{\begin{eqnarray}}
\newcommand{\ee}{\end{eqnarray}}
\begin{document}
\pagestyle{plain}
\title{Weighted Laplacians, cocycles and recursion relations}
\author{Kirill Krasnov${}^{1}$ and Carlos Scarinci${}^{2}$  \\ \\ \it{${}^1$ \, School of Mathematical Sciences, University of Nottingham}\\ \it{University Park, Nottingham, NG7 2RD, UK} \and 
\it{${}^2$ \, Department Mathematik, FAU Erlangen-N\"urnberg}  \\ \it{Cauerstrasse 11, 91058 Erlangen, Germany}}
\date{October 2013}
 \maketitle
 \begin{abstract}\noindent Hodge's formula represents the gravitational MHV amplitude as the determinant of a minor of a certain matrix. When expanded, this determinant becomes a sum over weighted trees, which is the form of the MHV formula first obtained by Bern, Dixon, Perelstein, Rozowsky and rediscovered by Nguyen, Spradlin, Volovich and Wen. The gravity MHV amplitude satisfies the Britto, Cachazo, Feng and Witten recursion relation. The main building block of the MHV amplitude, the so-called half-soft function, satisfies a different, Berends-Giele-type recursion relation. We show that all these facts are illustrations to a more general story. 
 
 We consider a weighted Laplacian for a complete graph of $n$ vertices. The matrix tree theorem states that its diagonal minor determinants are all equal and given by a sum over spanning trees. We show that, for any choice of a cocycle on the graph, the minor determinants satisfy a Berends-Giele as well as Britto-Cachazo-Feng-Witten type recursion relation. Our proofs are purely combinatorial. 
\end{abstract}

\section{Introduction and the main theorems}

The purpose of this note is to point out that there exists a certain generalization of recursion relations that are of central importance in the subject of (gravitational) scattering amplitudes. Over the past few years it has been realized that these amplitudes are closely related to matrices that have the property that the sum of elements in any row or column is zero. This relation is made particularly strong by the recent work \cite{Cachazo:2013hca} that builds scattering amplitudes of a variety of theories in arbitrary number of dimensions from determinants of minors of such matrices. Here we point out that certain recursion relations known to exist for the gravitational scattering amplitudes generalize to the context of such matrices. 

Our arguments are purely combinatorial. Thus, while our examples come from the gravitational perturbation theory and can be motivated by studying Feynman diagrams, there are no physics considerations in this paper. It is possible that the recursion relations we point out are known to combinatorists, even though we were unable to find them in the literature. If this is the case, we hope that our paper will be at least useful as bringing these generalizations to the attention of the scattering amplitudes community. 

With our paper being mainly addressed to the scattering amplitudes community, we start with a review of the main objects as they appear in this context. Readers interested just in the combinatorial statements can find these at the end of this Introduction. 

\subsection{Recursion relations for scattering amplitudes}

Tree-level scattering amplitudes can be efficiently computed using recursion relations. There are two main types of these. First, there are Berends-Giele \cite{Berends:1987me} recursions for currents: amplitudes for $n$ on-shell and one off-shell particle. The $n+1$ particle current is then constructible from the known currents with up to $n$ particles. The Berends-Giele recursion relations can become particularly simple when all on-shell particles are of the same type (e.g. have the same helicity). For instance, for the case of QCD, only Feynman diagrams with trivalent vertices contribute to the $n$-current of particles with the same helicity. The resulting Berends-Giele recursion can be easily solved, and the formula for the $n$-current contains all the main features of the famous Parke-Taylor MHV amplitude formula \cite{Parke:1986gb}. In a sense, one can say that the formula for all same helicity particles $n$-current is more fundamental than the MHV amplitude formula in that the latter can be guessed if the former is known. In the case of QCD the MHV amplitude formula itself can be obtained by solving the Berends-Giele recursion relation for the all but one particle of the same helicity, see \cite{Berends:1987me}. 

For gravity, the current for $n$ gravitons of the same helicity was first computed using Berends-Giele recursions in \cite{Bern:1998sv}. Abusing the terminology somewhat, we shall refer to this as the MHV gravitational current. In \cite{Bern:1998sv}, the Berends-Giele recursion relation for the MHV gravitational current was made simple and solvable by rewriting the trivalent gravitational vertex as a square of the gauge theory one. In contrast to the situation in QCD, for gravity the recursion for the all but one same helicity current appears to be too complicated to solve, and the gravitational MHV amplitude was never directly obtained via this route. However, the obtained in \cite{Bern:1998sv} MHV current formula leads to a simple and natural guess for the MHV amplitude itself. This guess was later shown to be correct, see below. So, in a sense the MHV current "knows" about the MHV amplitude in that the latter can be guessed from the formed using the soft and collinear factorisation properties. The fact that the "MHV current" contains so much information about the MHV amplitude justifies this terminology. 

The second type of recursion relations works with on-shell amplitudes only. This is the famous Britto-Cachazo-Feng-Witten (BCFW) recursion \cite{Britto:2005fq}. It glues together on-shell amplitudes to form on-shell amplitudes with more particles. For the MHV amplitudes (in both gauge theory and gravity) it takes a particularly simple form in that it only needs the amplitude with $n$ particles to construct the $n+1$ amplitude. This is in contrast to the Berends-Giele type recursion that needs all currents with up to $n$ particles to get the $n+1$ particle current. It is not always easy, however, to obtain a simple closed form of the solution to this recursion. Thus, in the case of gravity its main usage is to prove a formula guessed by some other means rather than to suggest a particular way of writing the solution. It can also be implemented on the computer, and in this way determines any scattering amplitude.

The first known closed form of the gravitational MHV amplitude \cite{Berends:1988zp} was obtained by a string-theoretic argument relating gravity to a square of the gauge theory. The first direct self-contained proof was given by \cite{Mason:2008jy}. In both of these works, the obtained MHV formula is rather complicated, hiding, in particular, the symmetries of the result. A completely different line of thought was that in the already mentioned work \cite{Bern:1998sv}, where the known expression for the gravitational MHV current allowed the authors to propose a natural guess for the MHV amplitude formula. The formula in \cite{Bern:1998sv} is arguably the most beautiful expression for the MHV amplitude available. More recently it was rediscovered and proved in \cite{Nguyen:2009jk}, and served as a basis for yet another interpretation \cite{Hodges:2012ym} of the MHV amplitude in terms of a certain determinant. 

\subsection{The MHV current and Berends-Giele recursion}

We shall define the MHV current as a purely combinatorial object, forgetting about the fact that this object arises from gravity Feynman diagrams. We, however, recall that this object is the main building block of the amplitude for $n$ on-shell gravitons of the same helicity, and one off-shell graviton. Gravitons being massless, their 4-momentum $p^\mu=p^{AA'}$, where $A,A'$ are the spinor indices, is a product of two spinors $p^{AA'}=p^A p^{A'}$. We will use the usual in the literature notation $p_i^A\equiv i^A, p_i^{A'}\equiv i^{A'}$. We denote the contraction of two unprimed spinors by a round bracket $i^A j_A \equiv (ij)$, and that of two primed spinors by a square bracket $i_{A'} j^{A'}\equiv [ij]$. The object that we shall refer to as the MHV current depends on the spinors $i^A, i^{A'}$ for the $n$ on-shell particles, as well as on a reference spinor $q^A$. There is no dependence on the complex conjugate $q^{A'}$ spinor. 

Let us define the 1-current via
$$A(1)=\frac{1}{(1q)^4}.$$
This is a complex number that depends on a single unprimed spinor $1^A$, as well as on the reference spinor $q^A$. The higher currents can be successively be built up using the Berends-Giele recursion relation
\be\label{BG}
\left(\sum_{i,j\in\cK}(ij)[ij]\right) A(\cK)=\sum_{\genfrac{}{}{0cm}{3}{\cI\subsetneq\cK}{\cJ=\cK\backslash\cI}}A(\cI) A(\cJ) \Big(\sum_{\genfrac{}{}{0cm}{3}{i\in\cI}{j\in\cJ}}(iq)(jq)[ij]\Big)^2.
\ee
The quantities appearing here are as follows. First, $\cK$ is a set of momenta on which the current depends. The sum on the right-hand-side is taken over all splittings of the set $\cK$ into two subsets $\cI,\cJ$. Thus, the sum starts with terms where $\cI$ is a set of one element. Then a sum over all possibilities for this element (i.e. all elements of $\cK$) is taken. Then it continues for $\cI$ being a set of two elements, etc. The last terms in the sum are where $\cI$ is a set of all but one element, which is then $\cJ$, and the sum is taken over this element. We note that many terms appear in the sum twice, but this is convenient for our purposes. We also note that the sum on the left-hand-side is over all elements $i,j\in\cK$, and so there is another double counting. This cancels the double counting occurring in the sum over the sets $\cI, \cJ$. It is clear that the relation (\ref{BG}) has its origin in some Feynman diagrams, as the left-hand-side (which can be put into the denominator on the right-hand-side) is just the total sum of momenta in $\cK$ squared. We, however, will refrain from explaining the Feynman rules that lead to this current. They can be found in either \cite{Bern:1998sv} in the metric formalism, or in \cite{Delfino:2012aj} in the "pure connection" formalism. In both of these formalisms the current is given by $A(\cK)$ satisfying (\ref{BG}), after stripping some inessential factors. 

To see how the recursion (\ref{BG}) determines the current let us work out some first few cases. The current $A(1,2)$ is particularly simple. In this case there are just two terms in the sum, $\cI=1,\cJ=2$ and $\cI=2,\cJ=1$, which both give the same result. The arising factor of 2 is then cancelled by a factor of 2 from the denominator and one obtains
\be\label{12}
A(1,2)=\frac{1}{(1q)^2(2q)^2}\frac{[12]}{(12)}.
\ee
To get a feel for how the recursion relation works we will list another example
$$A(1,2,3)=\frac{1}{(1q)^2(3q)^2}\frac{[12][23]}{(12)(23)}+\frac{1}{(1q)^2(2q)^2}\frac{[23][31]}{(23)(31)}+\frac{1}{(2q)^2(3q)^2}\frac{[31][12]}{(31)(12)}.$$
It is now not hard to guess the general formula. We have
\be\label{current}
A(\cK)=\prod_{i\in\cK}\frac{1}{(iq)^4}\sum_{T\in\cT(\cK)}\prod_{\langle jk\rangle\in E(T)}\frac{[jk]}{(jk)}(jq)^2(kq)^2.
\ee
Here the product in front is just that over all elements of $\cK$, then the sum is taken over all trees $T$ from the set $\cT(\cK)$ of trees with elements of $\cK$ as vertices. The notation $E(T)$ stands for the set of edges of a tree $T$. The formula (\ref{current}) can be easily tested for small sets $\cK$ to give the correct result. Thus, when $\cK=\{1,2\}$ is a set of two elements, the only tree is one connecting $1$ with $2$. So, the sum in this case consists of one term, which is (\ref{12}). The general formula (\ref{current}) can then be proved, see \cite{Bern:1998sv}, to satisfy the recursion (\ref{BG}), and this establishes it as a correct closed form expression for the current. We give a combinatorial proof in the main text, after this recursion is generalised to the weighted Laplacians context. 

\subsection{Interpretation in terms of a determinant}

It is not hard to see that the quantity (\ref{current}) is related to a certain matrix determinant. As far as we know this interpretation (of a related quantity) has first appeared in writing in \cite{Feng:2012sy}. 

Consider a complete graph whose vertices are elements from our set $\cK$. This is a graph where every vertex is connected to every other vertex (by exactly one edge). We now define a weighted Laplacian matrix for our complete graph, by associating 
\be
w_{ij} = \left\{ \lower2ex\vbox{ \hbox{$-\frac{[ij]}{(ij)}(iq)^2(jq)^2 ,\quad i\not=j.$}\hbox{ $\sum_{k\not=i} \frac{[ik]}{(ik)}(iq)^2(kq)^2, \quad i=j,$}} \right.
\ee
Then the determinant of the submatrix $W'$ of $W$ obtained by deleting $i$-th row and $i$-th column is independent of $i$ can can be denoted as $|W'|$. It is given by the following sum over trees
\be
|W'| = \sum_{T\in\cT(\cK)}\prod_{\langle ij\rangle\in E(T)} w_{ij}.
\ee
Looking at (\ref{current}) we see that a multiple of our current is expressible as such a matrix determinant:
\be\label{A-det}
A(\cK) \prod_{i\in\cK}(iq)^4 = |W'|.
\ee

\subsection{The half-soft function}

The half-soft functions were introduced in \cite{Bern:1998sv}, and are really just generalisations of the currents $A(\cK)$ to two different reference spinors. Unlike the currents, they appear to have no direct link to Feynman diagrams. However, they satisfy a recursion relation similar to (\ref{BG}). Their expression in terms of sum over trees is given by
\be\label{h}
h(x,\cK,y) = \prod_{i\in\cK}\frac{1}{(ix)^2(iy)^2}\sum_{T\in\cT(\cK)}\prod_{\langle jk\rangle\in E(T)}\frac{[jk]}{(jk)}(jx)(jy)(kx)(ky).
\ee
It is clear that
\be
h(q,\cK,q)=A(\cK).
\ee
The half-soft functions satisfy the following Berends-Giele-type recursion relation 
\be\label{BG-h}
\left(\sum_{i,j\in\cK}(ij)[ij]\right) h(x,\cK,y)=\sum_{\genfrac{}{}{0cm}{3}{\cI\subsetneq\cK}{\cJ=\cK\backslash\cI}}h(x,\cI,y) h(x,\cJ,y) \Big(\sum_{\genfrac{}{}{0cm}{3}{i\in\cI}{j\in\cJ}}(ix)(jx)[ij]\Big)\Big(\sum_{\genfrac{}{}{0cm}{3}{i\in\cI}{j\in\cJ}}(iy)(jy)[ij]\Big).
\ee
The half-soft functions can also be represented as the determinant of a minor of a matrix, see \cite{Feng:2012sy} for an explicit formula, and also below.

\subsection{MHV amplitudes}

A simple formula for the gravitational MHV amplitudes in terms of the half-soft function is possible. This first appeared as a guess in \cite{Bern:1998sv}, and then was rediscovered and proved in \cite{Nguyen:2009jk}. 

Let $1,2$ be the momenta of two positive helicity gravitons, and $\cK$ be the set of momenta of the negative helicity gravitons. Then 
\be
{\cal M}^{\rm MHV} = 2(\kappa/2)^{|\cK|} (12)^6 \, h(1,\cK,2),
\ee
where $\kappa^2=32\pi G$. Thus, in this formula the momentum spinors of the positive helicity gravitons are used as the references spinors of the half-soft function. 

\subsection{The soft factor}

Let us consider the limit as one of the momenta, say $1$, in the set $\cK$ goes to zero. As we can see from the formula (\ref{h}), there are occurrences of $(1x),(1y)$ in the denominator. On the other hand, the ratios $[1k]/(1k)$, being ratios, do not blow up in the limit. Concentrating on the quantities $(1x),(1y)$, one can easily see that the total power of $(1x)$ in each term in the sum is $(m_1-2)$, where $m_1$ is the number of edges that are connected to $1$ in the tree. So, we see that the terms with just one edge connected to $1$ are singular, because their dependence on $1\to0$ is $1/(1x)$. The terms with two or more edges are, on the other hand, non-singular, and can be dropped in the limit as less important as compared to the singular terms. So, we see that in the limit we should just keep in (\ref{h}) the trees where $1$ is connected to the rest of the tree just by one edge. It is then easy to see that
\be\label{soft}
h(x,\cK,y) \Big|_{1\to 0} \to S_1(x,\cK,y)  h(x,\cK\backslash \{1\},y),
\ee
where the soft factor is given by
\be
S_1(x,\cK,y) = \sum_{l\in\cK\backslash\{1\}}\frac{[1l]}{(1l)}\frac{(lx)(ly)}{(1x)(1y)}.
\ee
The $l$ in sum in the soft factor has the interpretation of the vertex in $\cK\backslash \{1\}$ tree to which the vertex $1$ is attached. 

\subsection{The BCFW recursion}

The MHV amplitudes, being on-shell amplitudes, satisfy the BCFW \cite{Britto:2005fq} recursion relation. There are several types of this recursion, depending on which combination of the helicities is used for the BCFW analytic continuation. We state it in a version in which the resulting recursion takes the form of the "inverse-soft" reconstruction, see e.g. \cite{BoucherVeronneau:2011nm,Dunbar:2012aj}. This is also the form used by Hodges \cite{Hodges:2012ym}. 

The BCFW recursion for the MHV amplitude, after dividing by the factor $(12)^6$, implies such a recursion for the half-soft function. Let us define
\be\label{shift}
\hat l'=l'+\frac{(1x)}{(lx)}1',
\ee
which is a familiar BCFW-type shift. Then the $n$-th half-soft function is constructed from the $(n-1)$-th by taking a sum over all possible vertices in $\cK\backslash \{1\}$ to which the new vertex $1$ can be attached:
\be\label{BCFW}
h(x,\cK,y)=\sum_{l\in\cK\backslash\{1\}}\frac{[1l]}{(1l)}\frac{(lx)(ly)}{(1x)(1y)}h(x,\cK^{\hat l}\backslash\{1\},y).
\ee
When $1\to 0$ the shift disappears $\hat{l}'\to l'$, and we automatically recover the correct soft behaviour (\ref{soft}). This is why the formula of the type (\ref{BCFW}) can be referred to as the inverse-soft recursion. The recursion (\ref{BCFW}) also holds with $x$ replaced by $y$ in (\ref{shift}). 

\subsection{Matrix tree theorem}

We now generalise both of the above recursions to the context of weighted Laplacians. Let us start with the matrix tree theorem. 

Let $\cK=\{1,...,n\}$ denote a set of vertices and $G(\cK)$ be the complete graph on $\cK$ with each vertex connected all the others by a single edge. We denote the edge from $i$ to $j$ by $\langle ij\rangle$. A spanning tree $\gamma$ on $\cK$ (or rather $G(\cK)$) is a subgraph of $G(\cK)$ with no loops which passes through all the vertices. Let $\cT^\cK=\cT(G(\cK))$ denote the set of spanning trees on $G(\cK)$. 

Let's assign to each edge $\langle ij\rangle$ of $G(\cK)$ a weight $w(\langle ij\rangle)=w_{ij}$. We then define the weighted Laplacian of $(\cK,w)$ as the matrix $\Delta^{(\cK,w)}$ whose entries are given by
$$\Delta^{(\cK,w)}_{ij}=\begin{cases}-w_{ij},\quad i\neq j \cr \sum_{k\neq i}w_{ik},\quad i=j\end{cases}.$$

The matrix tree theorem states that the (diagonal) minor determinants of $\Delta^{(\cK,w)}$ are all equal and given by
\be\label{MT}
|\Delta^{(\cK,w)}|=\sum_{\gamma\in\cT^\cK}\prod_{\langle ij\rangle\in\gamma}w_{ij}=\sum_{\gamma\in\cT^\cK}w(\gamma).
\ee
For a set $|\cK|=1$ consisting of a single element we define $|\Delta^{(\cK,w)}|:=1$.

\subsection{Cocycle}

We now introduce an element crucial for the construction that follows. Let $C_{ij}$ be a cocycle on $G(\cK)$, i.e. a function on edges of the complete graph of $\cK$ with the properties
\be
C_{ji}=-C_{ij}, \qquad C_{ij}+C_{jk}+C_{ki}=0.
\ee
The main result presented in this paper is that, for any choice of such cocycle, there is a version of the Berends-Giele as well as BCFW recursion formulas for $|\Delta^{(\cK,w)}|$. 

\subsection{Recursions for weighed Laplacians}

Our first theorem is a Berends-Giele type recursion relation.
\begin{theorem} For any choice of a pair $C_{ij},\tilde{C}_{ij}$ of cocycles  on $G(\cK)$, the weighted Laplacian $\Delta^{(\cK,w)}$ satisfies the recursion relation
\be\label{rec-BG}
|\Delta^{(\cK,w)}|\Big(\sum_{i,j\in\cK}C_{ij} \tilde{C}_{ij} w_{ij}\Big)=\sum_{\cI\sqcup\cJ=\cK}|\Delta^{(\cI,w)}||\Delta^{(\cJ,w)}|\Big(\sum_{i\in\cI,j\in\cJ}C_{ij}w_{ij}\Big)\Big(\sum_{k\in\cI,l\in\cJ}\tilde{C}_{kl}w_{kl}\Big).
\ee
\end{theorem}

The second theorem is a statement of a BCFW-type recursion.
\begin{theorem} For any cocycle $\alpha_{ij}$ on $G(\cK)$, the weighted Laplacian $\Delta^{(\cK,w)}$ satisfies the recursion relation 
\be\label{rec-BCFW}
|\Delta^{(\cK,w)}|=\sum_{l\in\cK\backslash\{1\}}w_{1l}\, |\Delta^{(\cK^{\hat l}\backslash\{1\},w)}|
\ee
where the weights are shifted by the cocycle as 
$$w_{i\hat l}:=w_{il}+\frac{C_{i1}}{C_{il}}w_{i1}.$$
\end{theorem}

\subsection{The case of gravity}

In the case of gravitational MHV amplitudes and the corresponding recursion relations, the pair of cocycle from (\ref{rec-BG}) is given by
\be\label{ab}
C_{ij} = \frac{(ij)}{(ix)(jx)}, \qquad \tilde{C}_{ij} = \frac{(ij)}{(iy)(jy)}.
\ee
Both are obviously anti-symmetric $C_{ji}=-C_{ij}$, and also "exact", i.e. can be represented as a difference $C_{ij} = f(i)-f(j)$, where the function $f:\cK\to{\mathbb C}$ can be extracted by multiplying the numerator and denominator of the cocycle by $(px)$, where $p$ is some other reference spinor, and using the Shouten identity. This gives $f(i) = (ip)/(ix)$. This representation, in particular, proves the cocycle property of $C_{ij}$. With this choice of the cocycle, and defining
\be
w_{ij} = \frac{[ij]}{(ij)} (ix)(iy) (jx)(jy),
\ee
recursion (\ref{rec-BG}), after being multiplied by $\prod_{i\in\cK} (ix)^2(iy)^2$, reduces to (\ref{BG-h}). The recursion (\ref{BG}) is obtained by setting $x=y=q$, which makes the cocycles equal $C_{ij}=\tilde{C}_{ij}$ and converts $h(x,\cK,y)$ into $A(\cK)$. 

To obtain (\ref{BCFW}) with the shift (\ref{shift}) one chooses $C_{ij}$ to be the second cocycle in (\ref{ab}).  When multiplied by $\prod_{i\in\cK} (ix)^2(iy)^2$ the recursion (\ref{rec-BCFW}) reduces to (\ref{BCFW}). If one instead chooses $C_{ij}$ to be the first cocycle in (\ref{ab}) one obtains the recursion (\ref{BCFW}) where $x$ is replaced by $y$ in the shift (\ref{shift}). 

\subsection{Acknowledgement} KK is grateful to F. Cachazo and D. Skinner for a discussion. KK was supported by an ERC grant 277570-DIGT, and partially by a fellowship from the Alexander von Humboldt foundation, Germany.

\section{Proof of the Berends-Giele-type recursion}

Let us introduce some additional terminology. A rooted tree is a tree with a preferred vertex. A forest is a collection of disjoint trees and a rooted forest is a collection of disjoint rooted trees. Let $\cF^\cK_{a_1a_2...a_r}$ denote the set of all spanning rooted forests on $\cK$ with roots $a_1,a_2,...,a_r\in\cK$.

We prove (\ref{rec-BG}) by explicitly substituting (\ref{MT}) into the right-hand-side of (\ref{rec-BG}), and showing that the terms organise themselves into a sum over weighed trees as given on the left-hand-side. 

Thus, we have 
\be\nonumber
\sum_{\cI\sqcup\cJ=\cK}|\Delta^{(\cI,w)}||\Delta^{(\cJ,w)}|\Big(\sum_{a\in\cI,b\in\cJ}C_{ab}w_{ab}\Big)\Big(\sum_{c\in\cI,d\in\cJ}\tilde{C}_{cd}w_{cd}\Big)
\\ \nonumber
=\sum_{\cI\sqcup\cJ=\cK}\sum_{\gamma\in\cT^\cI,\delta\in\cT^\cJ}\prod_{\langle ij\rangle\in\gamma,\delta} w_{ij} \sum_{a\in\cI,b\in\cJ}C_{ab}w_{ab} \sum_{c\in\cI,d\in\cJ}\tilde{C}_{cd} w_{cd}
\\ \nonumber
=\sum_{\cI\sqcup\cJ=\cK}\sum_{a\in\cI,b\in\cJ}\sum_{\gamma\in\cT^\cI,\delta\in\cT^\cJ}\prod_{\langle ij\rangle\in\gamma,\delta,\langle ab\rangle}w_{ij}\Big(\sum_{c\in\cI,d\in\cJ}C_{ab}\tilde{C}_{cd} w_{cd}\Big)
\\ \nonumber
=2\sum_{a,b\in\cK}\sum_{f_{ab}\in\cF^\cK_{ab}}\prod_{\langle ij\rangle\in f_{ab},\langle ab\rangle}w_{ij}\Big(\sum_{c\in\cK^a_{f},d\in\cK^b_{f}}C_{ab}\tilde{C}_{cd}w_{cd}\Big)
\\ \nonumber
=\sum_{\gamma\in\cT^\cK}\prod_{\langle ij\rangle\in\gamma}w_{ij}\sum_{\langle ab\rangle\in\gamma}\Big(\sum_{c\in\cK^a_\gamma,d\in\cK^b_\gamma}C_{ab}\tilde{C}_{cd}w_{cd}\Big).
\ee
In the second line we simply substituted the formula (\ref{MT}) for each of the quantities $|\Delta^{(\cI,w)}|, |\Delta^{(\cJ,w)}|$, as well as expanded the product of two brackets in the first line. In the third line we absorb into the product the weight factor of $\langle ab\rangle$. We also reorder the sum by considering the sum over $a\in\cI,b\in\cJ$ before the sum over spanning trees of $\cI$ and $\cJ$. Now comes the crux of the calculation. Each element of the sum in the third line can be depicted by a spanning tree composed by a rooted spanning forest of two trees and an edge connecting their roots. This is equivalent to a tree with one preferred edge. The value of this element is then given by the weight of this tree (equal to the weight of the forest times the weight of the edge connecting its roots) times a factor (in parenthesis) depending on the tree and its preferred edge. Since we have a sum over all decompositions of $\cK$ into disjoint $\cI,\cJ$ and a sum over all elements of these sets the sum in the third line ranges through all rooted forests of $\cK$. Note that since we have a non-ordered sum over decompositions of $\cK$ each rooted forest appears twice in the sum. This explains the fourth line. To go to the last line we simply reorder the sum by putting into evidence the weight of each spanning tree. Thus we get a sum over spanning trees of $\cK$ of the weight of the tree times the sum over its edges of the factor in parenthesis. The factor of 2 is now absorbed into the sum over edges since we need to consider both $\langle ab\rangle$ and $\langle ba\rangle$. The sum in the brackets in the last line is over elements $c$ of the set $\cK^a_\gamma$, and $d$ of set $\cK^b_\gamma$, where the sets $\cK^a_\gamma, \cK^b_\gamma: \mathcal{K}=\mathcal{K}^a_\gamma\sqcup\mathcal{K}^b_\gamma$ are those though which the graph $\gamma$ with the edge $\langle ab\rangle$ removed passes. We also have $a\in\mathcal{K}^a_\gamma$,$b\in\mathcal{K}^b_\gamma$.

To finish the proof we take a closer look at the expression
\be\label{bg-1}
\sum_{\langle ab\rangle\in\gamma}\Big(\sum_{c\in\cK^a_\gamma,d\in\cK^b_\gamma}C_{ab}\tilde{C}_{cd}w_{cd}\Big)
\ee
and show it equals
$$\Big(\sum_{c,d\in\cK}C_{cd}\tilde{C}_{cd} w_{cd}\Big).$$
To evaluate (\ref{bg-1}) we reorder the sum, first fixing $c,d$. Then the sum over edges $\langle ab\rangle$ is over those forming the direct path from $c$ to $d$ along $\gamma$
\be\nonumber
\sum_{\langle ab\rangle\in\gamma}\sum_{c\in\mathcal{K}^a_{\gamma},d\in\mathcal{K}^b_{\gamma}}C_{ab}\tilde{C}_{cd}w_{cd} =\sum_{c,d\in\mathcal{K}}\tilde{C}_{cd}w_{cd}\sum_{\langle ab\rangle\in\gamma_{cd}}C_{ab}=\sum_{c,d\in\mathcal{K}}C_{cd}\tilde{C}_{cd}w_{cd},
\ee
where to get the last equality we have used the cocycle property
$$\sum_{\langle ab\rangle\in\gamma_{cd}}C_{ab}=C_{cd}.$$
This finishes the proof. 

\section{Proof of the BCFW-type recursion}

The proof of (\ref{rec-BCFW}) is much harder. The idea of the proof is to use the already proven recursion relation (\ref{rec-BG}), as well as the following, still unproven, identity
\be\label{ident}
|\Delta^{(\cK,w)}| \sum_{l\in\cK}w_{pl}C_{pl}=\sum_{l\in\cK}w_{pl}C_{pl}|\Delta^{(\cK^{\hat l},w)}|,
\ee
where $p$ is the element of the set directing the shift
\be
w_{k\hat{l}}=w_{kl} + \frac{C_{kp}}{C_{kl}} w_{kp},
\ee
and prove the recursion (\ref{rec-BCFW}) by induction.

\subsection{Proof of the identity}

Let's start with the identity. Here we shall use a diagrammatic notations to perform the purely combinatorial part of the computation. We denote the shifted Laplacian $|\Delta^{(\cK^{\hat l},w)}|$ diagrammatically as a sum over trees $\vec\tau$ with an additional oriented edge on top of each edge connected to the shifted vertex.

For example we draw the Laplacians with two and three vertices as
\begin{figure}[H]
\centering
\begin{tikzpicture}[scale=0.5]
\node at (-3.0,0.5) {$|\Delta^{(\{1,2\},w)}|=\;$};
\draw (-0.75,0.5) -- (0.75,0.5);
\end{tikzpicture}
\\
\begin{tikzpicture}[scale=0.5]
\node at (-3.0,0.5) {$|\Delta^{(\{1,2,3\},w)}|=\;$};
\draw (-0.75,0) -- (0,1) -- (0.75,0);
\draw  (3,1) --(3-0.75,0) -- (3.75,0);
\draw (6-0.75,0) -- (6.75,0) -- (6,1) ;
\node at (1.5,0.5) {$+$};
\node at (4.5,0.5) {$+$};
\end{tikzpicture}
\end{figure}
\noindent where we understand that each edge is weighted and the value of each term is given by the product of the weight of its edges. When we shift the momenta we draw
\begin{figure}[H]
\centering
\begin{tikzpicture}[scale=0.5]
\node at (-3,0.5) {$|\Delta^{(\{\hat 1,2\},w)}|=\;$};
\draw (-0.75,0.5) -- (0.75,0.5);
\draw [->](-0.75,0.5+0.1) -- (0.75,0.5+0.1);
\node at (-0.5+7-2.5,0.5) {$|\Delta^{(\{1,\hat 2\},w)}|=\;$};
\draw (7-0.75,0.5) -- (7.75,0.5);
\draw [<-](7-0.75,0.5+0.1) -- (7.75,0.5+0.1);
\end{tikzpicture}
\\
\begin{tikzpicture}[scale=0.5]
\node at (-3.4,0.5) {$|\Delta^{(\{\hat 1,2,3\},w)}|=\;$};
\draw (-0.75,0) -- (0,1) -- (0.75,0);
\draw [->] (0-0.1,1+0.1) -- (-0.75-0.1,0+0.1);
\draw [->] (0+0.1,1+0.1) -- (0.75+0.1,0+0.1);
\draw  (3,1) --(3-0.75,0) -- (3.75,0);
\draw [->] (3-0.1,1+0.1) --(3-0.75-0.1,0+0.1);
\draw (6-0.75,0) -- (6.75,0) -- (6,1) ;
\draw [<-] (6.75+0.1,0+0.1) -- (6+0.1,1+0.1);
\node at (1.5,0.5) {$+$};
\node at (4.5,0.5) {$+$};
\end{tikzpicture}
\\
\begin{tikzpicture}[scale=0.5]
\node at (-0.5-2.9,0.5) {$|\Delta^{(\{1,\hat 2,3\},w)}|=\;$};
\draw (-0.75,0) -- (0,1) -- (0.75,0);
\draw [<-] (0-0.1,1+0.1) -- (-0.75-0.1,0+0.1);
\draw  (3,1) --(3-0.75,0) -- (3.75,0);
\draw [<-] (3-0.1,1+0.1) --(3-0.75-0.1,0+0.1);
\draw [->] (3-0.75,0-0.1)-- (3.75,0-0.1);
\draw (6-0.75,0) -- (6.75,0) -- (6,1) ;
\draw [->] (6-0.75,0-0.1)-- (6.75,0-0.1);
\node at (1.5,0.5) {$+$};
\node at (4.5,0.5) {$+$};
\end{tikzpicture}
\\
\begin{tikzpicture}[scale=0.5]
\node at (-0.5-2.9,0.5) {$|\Delta^{(\{1,2,\hat 3\},w)}|=\;$};
\draw (-0.75,0) -- (0,1) -- (0.75,0);
\draw [<-] (0+0.1,1+0.1) -- (0.75+0.1,0+0.1);
\draw  (3,1) --(3-0.75,0) -- (3.75,0);
\draw [<-] (3-0.75,0-0.1)-- (3.75,0-0.1);
\draw (6-0.75,0) -- (6.75,0) -- (6,1) ;
\draw [->] (6.75+0.1,0+0.1) -- (6+0.1,1+0.1);
\draw [<-] (6-0.75,0-0.1)-- (6.75,0-0.1);
\node at (1.5,0.5) {$+$};
\node at (4.5,0.5) {$+$};
\end{tikzpicture}
\end{figure}
\noindent where we interpret the arrows as being added to the corresponding edges. Therefore we may use distributivity to draw
\begin{figure}[H]
\centering
\begin{tikzpicture}[scale=0.5]
\draw (-0.75,0) -- (0,1) -- (0.75,0);
\draw [<-] (0+0.1,1+0.1) -- (0.75+0.1,0+0.1);
\node at (1.5,0.5) {$=$};
\draw (3-0.75,0) -- (3,1) -- (3.75,0);
\draw (6-0.75,0) -- (6,1);
\draw [->] (6.75,0) -- (6,1);
\node at (4.5,0.5) {$+$};
\end{tikzpicture}
\\
\begin{tikzpicture}[scale=0.5]
\draw (-0.75,0) -- (0.75,0) -- (0,1) ;
\draw [->] (0.75+0.1,0+0.1) -- (0.1,1+0.1);
\draw [<-] (-0.75,0-0.1)-- (0.75,0-0.1);
\node at (1.5,0.5) {$=$};
\draw (3-0.75,0) -- (3.75,0) -- (3,1) ;
\node at (4.5,0.5) {$+$};
\draw (6-0.75,0) -- (6.75,0);
\draw [->] (6.75,0) -- (6,1);
\node at (7.5,0.5) {$+$};
\draw (9.75,0) -- (9,1) ;
\draw [<-] (9-0.75,0)-- (9.75,0);
\node at (10.5,0.5) {$+$};
\draw [->] (12.75,0) -- (12,1) ;
\draw [<-] (12-0.75,0)-- (12.75,0);
\end{tikzpicture}
\end{figure}

Drawing all terms in the identity in this manner, we obtain terms without any arrows, terms with a single arrow, a pair arrows, a triple of arrows and so on, until all edges have arrows. The terms with no arrows cancel exactly the left-hand-side of the identity and the remaining terms cancel by themselves order by order in the number of arrows. For example the identity with two points is 
\begin{figure}[H]
\centering
\begin{tikzpicture}[scale=0.5]
\draw (-0.75,0.5) -- (0.75,0.5);
\node at (4.5,0.5) {$\Big(w_{p1}C_{p1}+w_{p2}C_{p2}\Big)=$};
\draw (9-0.75,0.5) -- (9.75,0.5);
\draw [->](9-0.75,0.5+0.1) -- (9.75,0.5+0.1);
\node at (7+4.5,0.5) {$w_{p1}C_{p1}+$};
\draw (14-0.75,0.5) -- (14.75,0.5);
\draw [<-](14-0.75,0.5+0.1) -- (14.75,0.5+0.1);
\node at (11.6+4.5,0.5) {$w_{p2}C_{p2}$};
\end{tikzpicture}
\\
\begin{tikzpicture}[scale=0.5]
\node at (-1.5,0.5) {$=\Big($};
\draw (-0.75,0.5) -- (0.75,0.5);
\draw [->] (3-0.75,0.5) -- (3.75,0.5);
\node at (1.5,0.5) {$+$};
\node at (5.3,0.5) {$\Big)w_{p1}C_{p1}$};
\node at (7.2,0.5) {$+\Big($};
\draw (8.5-0.75,0.5) -- (8.5+0.75,0.5);
\draw [<-] (11.5-0.75,0.5) -- (11.5+0.75,0.5);
\node at (10,0.5) {$+$};
\node at (13.8,0.5) {$\Big)w_{p2}C_{p2}$};
\end{tikzpicture}
\end{figure}
\noindent which is equivalent to the single-arrow identity
\begin{figure}[H]
\centering
\begin{tikzpicture}[scale=0.5]
\node at (-1.5,0.5) {$0=$};
\draw [->] (-0.75,0.5) -- (0.75,0.5);
\node at (2.5,0.5) {$w_{p1}C_{p1}\;+$};
\draw [<-] (5-0.75,0.5) -- (5+0.75,0.5);
\node at (7,0.5) {$w_{p2}C_{p2}$};
\end{tikzpicture}
\end{figure}
\noindent This will be proven later. Then, the three point identity reads
\begin{figure}[H]
\centering
\begin{tikzpicture}[scale=0.5]
\node at (-1.2,0.5) {$\Big($};
\draw (-0.75,0) -- (0,1) -- (0.75,0);
\draw  (3,1) --(3-0.75,0) -- (3.75,0);
\draw (6-0.75,0) -- (6.75,0) -- (6,1) ;
\node at (1.5,0.5) {$+$};
\node at (4.5,0.5) {$+$};
\node at (11.8,0.5) {$\Big)\Big(w_{p1}C_{p1}+w_{p2}C_{p2}+w_{p3}C_{p3}\Big)$};
\end{tikzpicture}
\\ \vspace{0.2cm}
\begin{tikzpicture}[scale=0.5]
\node at (-1.5,0.5) {$=\Big($};
\draw (-0.75,0) -- (0,1) -- (0.75,0);
\draw [->] (0-0.1,1+0.1) -- (-0.75-0.1,0+0.1);
\draw [->] (0+0.1,1+0.1) -- (0.75+0.1,0+0.1);
\draw  (3,1) --(3-0.75,0) -- (3.75,0);
\draw [->] (3-0.1,1+0.1) --(3-0.75-0.1,0+0.1);
\draw (6-0.75,0) -- (6.75,0) -- (6,1) ;
\draw [->] (6+0.1,1+0.1) -- (6.75+0.1,0+0.1);
\node at (1.5,0.5) {$+$};
\node at (4.5,0.5) {$+$};
\node at (8.3,0.5) {$\Big)w_{p1}C_{p1}$};
\node at (10.2,0.5) {$+\Big($};
\draw (11.5-0.75,0) -- (11.5,1) -- (11.5+0.75,0);
\draw [<-] (11.5-0.1,1+0.1) -- (11.5-0.75-0.1,0+0.1);
\draw  (14.5,1) --(14.5-0.75,0) -- (14.5+0.75,0);
\draw [<-] (14.5-0.1,1+0.1) --(14.5-0.75-0.1,0+0.1);
\draw [->] (14.5-0.75,0-0.1)-- (14.5+0.75,0-0.1);
\draw (17.5-0.75,0) -- (17.5+0.75,0) -- (17.5,1) ;
\draw [->] (17.5-0.75,0-0.1)-- (17.5+0.75,0-0.1);
\node at (13,0.5) {$+$};
\node at (16,0.5) {$+$};
\node at (19.8,0.5) {$\Big)w_{p2}C_{p2}$};
\end{tikzpicture}
\\ \vspace{0.2cm}
\begin{tikzpicture}[scale=0.5]
\node at (-1.5,0.5) {$+\Big($};
\draw (-0.75,0) -- (0,1) -- (0.75,0);
\draw [<-] (0+0.1,1+0.1) -- (0.75+0.1,0+0.1);
\draw  (3,1) --(3-0.75,0) -- (3.75,0);
\draw [<-] (3-0.75,0-0.1)-- (3.75,0-0.1);
\draw (6-0.75,0) -- (6.75,0) -- (6,1) ;
\draw [->] (6.75+0.1,0+0.1) -- (6+0.1,1+0.1);
\draw [<-] (6-0.75,0-0.1)-- (6.75,0-0.1);
\node at (1.5,0.5) {$+$};
\node at (4.5,0.5) {$+$};
\node at (8.3,0.5) {$\Big)w_{p3}C_{p3}$};
\end{tikzpicture}
\\ \vspace{0.2cm}
\begin{tikzpicture}[scale=0.5]
\node at (-1.5,0.5) {$=\Big($};
\draw (-0.75,0) -- (0,1) -- (0.75,0);
\draw (3-0.75,0) -- (3,1);
\draw [->] (3,1) -- (3.75,0);
\draw (6,1) -- (6.75,0);
\draw [->] (6,1) -- (6-0.75,0);
\draw [->] (9,1) -- (9-0.75,0);
\draw [->] (9,1) -- (9.75,0);
\draw  (12,1) --(12-0.75,0) -- (12.75,0);
\draw (15-0.75,0) -- (15.75,0);
\draw [->] (15,1) --(15-0.75,0);
\draw (18-0.75,0) -- (18.75,0) -- (18,1) ;
\draw (21-0.75,0) -- (21.75,0);
\draw [->] (21,1) -- (21.75,0);
\node at (1.5,0.5) {$+$};
\node at (4.5,0.5) {$+$};
\node at (7.5,0.5) {$+$};
\node at (10.5,0.5) {$+$};
\node at (13.5,0.5) {$+$};
\node at (16.5,0.5) {$+$};
\node at (19.5,0.5) {$+$};
\node at (23.5,0.5) {$\Big)w_{p1}C_{p1}$};
\end{tikzpicture}
\\ \vspace{0.2cm}
\begin{tikzpicture}[scale=0.5]
\node at (-1.5,0.5) {$+\Big($};
\draw (-0.75,0) -- (0,1) -- (0.75,0);
\draw [->] (3-0.75,0) -- (3,1);
\draw (3,1) -- (3.75,0);
\draw (6,1) --(6-0.75,0) -- (6.75,0);
\draw (9-0.75,0) -- (9.75,0);
\draw [<-] (9,1) --(9-0.75,0);
\draw [->](12-0.75,0) -- (12.75,0);
\draw (12,1) --(12-0.75,0);
\draw [<-] (15,1) --(15-0.75,0);
\draw [->](15-0.75,0) -- (15.75,0);
\draw (18-0.75,0) -- (18.75,0) -- (18,1) ;
\draw [->] (21-0.75,0) -- (21.75,0);
\draw (21.75,0) -- (21,1);
\node at (1.5,0.5) {$+$};
\node at (4.5,0.5) {$+$};
\node at (7.5,0.5) {$+$};
\node at (10.5,0.5) {$+$};
\node at (13.5,0.5) {$+$};
\node at (16.5,0.5) {$+$};
\node at (19.5,0.5) {$+$};
\node at (23.5,0.5) {$\Big)w_{p2}C_{p3}$};
\end{tikzpicture}
\\ \vspace{0.2cm}
\begin{tikzpicture}[scale=0.5]
\node at (-1.5,0.5) {$+\Big($};
\draw (-0.75,0) -- (0,1) -- (0.75,0);
\draw (3-0.75,0) -- (3,1);
\draw [<-](3,1) -- (3.75,0);
\draw (6,1) --(6-0.75,0) -- (6.75,0);
\draw [<-] (9-0.75,0) -- (9.75,0);
\draw (9,1) --(9-0.75,0);
\draw (12-0.75,0) -- (12.75,0) -- (12,1);
\draw [->](15-0.75,0) -- (15.75,0) -- (15,1);
\draw [<-](18-0.75,0) -- (18.75,0) -- (18,1);
\draw [<->](21-0.75,0) -- (21.75,0) -- (21,1);
\node at (1.5,0.5) {$+$};
\node at (4.5,0.5) {$+$};
\node at (7.5,0.5) {$+$};
\node at (10.5,0.5) {$+$};
\node at (13.5,0.5) {$+$};
\node at (16.5,0.5) {$+$};
\node at (19.5,0.5) {$+$};
\node at (23.5,0.5) {$\Big)w_{p3}C_{p3}$};
\end{tikzpicture}
\end{figure}
\noindent Clearly the terms with no arrows produce the left-hand-side of (\ref{ident}), and it is not hard to convince oneself that all single-arrow terms cancel by themselves using the single-arrow identity. Thus the two-point identity is equivalent to the two-arrow identity
\begin{figure}[H]
\centering
\begin{tikzpicture}[scale=0.5]
\node at (-1.5,0.5) {$0=$};
\draw [<->] (-0.75,0) -- (0,1) -- (0.75,0);
\draw [<->] (4.5,1) --(4.5-0.75,0) -- (4.5+0.75,0);
\draw [<->] (9-0.75,0) -- (9.75,0) -- (9,1) ;
\node at (3.5,0.5) {$+$};
\node at (8,0.5) {$+$};
\node at (2,0.5) {$w_{p1}C_{p1}$};
\node at (6.5,0.5) {$w_{p2}C_{p2}$};
\node at (11,0.5) {$w_{p3}C_{p3}$};
\end{tikzpicture}
\end{figure}

This pattern remains valid for identities with higher number of points. Thus the $n$-point identity is equivalent to a $n$-arrow identity
\begin{figure}[H]
\centering
\begin{tikzpicture}[scale=0.5]
\node at (-2.5,0.5) {$0=\displaystyle\sum_{l=1}^n$};
\draw [<-] (-1,-0.45) -- (0,0.5);
\draw [<-] (-1.25,0.5) -- (0,0.5);
\draw [<-] (-1,1.45) -- (0,0.5);
\draw [<-] (0,1.75) -- (0,0.5);
\draw [<-] (1,1.45) -- (0,0.5);
\draw [<-] (1.25,0.5) -- (0,0.5);
\node at (4.5+2.5,0.5) {$\displaystyle{w_{pl}C_{pl}=\sum_{l=1}^n\Big(\prod_{k\neq l}\frac{C_{kp}}{C_{kl}}w_{kp}\Big)w_{pl}C_{pl}}$};
\node at (0,0.2) { \rotatebox{61}{$\ddots$}};
\end{tikzpicture}
\end{figure}

Collecting some common factors we obtain
\be\label{bc-1}
\sum_{l=1}^n\Big(\prod_{k\neq l}\frac{C_{kp}}{C_{kl}}w_{kp}\Big)w_{pl}C_{pl}=\prod_{i=1}^nw_{ip}C_{ip}\sum_{l=1}^n\Big(\prod_{k\neq l}\frac{1}{C_{kl}}\Big)
\cr
=\prod_{i=1}^nw_{ip}C_{ip}\prod_{j<k}\frac{1}{C_{jk}}\sum_{l=1}^n\Big(\prod_{j<k}C_{jk}\prod_{m\neq l}\frac{1}{C_{ml}}\Big)
\cr
=(-1)^n\prod_{i=1}^nw_{ip}C_{ip}\prod_{j<k}\frac{1}{C_{jk}}\sum_{l=1}^n(-1)^l\prod_{\genfrac{}{}{0cm}{3}{j<k}{j,k\neq l}}C_{jk}.
\ee

Now, using the cocycle property $C_{jk}=C_{jr}-C_{kr}$ for some arbitrary reference vertex $r$, we identify (\ref{bc-1}) as a multiple of the (equal to zero) Vandermonde determinant
\be
\sum_{l=1}^n(-1)^l\prod_{\genfrac{}{}{0cm}{3}{j<k}{j,k\neq l}}C_{jk}=\det M_n
\ee
where the matrix $M_n$ is given by
\be
M_n=\left(\begin{matrix}1 & 1 & \cdots & 1
\cr
1 & 1 & \cdots & 1
\cr
C_{1r} & C_{2r} & \cdots & C_{nr}
\cr
\vdots & \vdots & \cdots & \vdots
\cr
C_{1r}^{n-2} & C_{2r}^{n-2} & \cdots & C_{nr}^{n-2}\end{matrix}\right).
\ee

\subsection{Proof by induction}

Now we turn to the induction part of the proof. The induction hypothesis is
$$|\Delta^{(\cI,w)}|=\sum_{l\in\cI\backslash\{1\}}w_{1l}|\Delta^{(\cI^{\hat l}\backslash\{1\},w)}|$$
for sets $\cI$ smaller than $\cK$ to prove the relation for $\cK$. We also note that the recursion (\ref{rec-BCFW}) obviously works for constructing $|\Delta^{(\cK,w)}|$ for $|\cK|=2$ from this object for $|\cK|=1$. 

Therefore we start by substituting the expression for $|\Delta^{(\cI,w)}|$ above into the already proved recursion relation (\ref{rec-BG}) with $C_{ij}=\tilde{C}_{ij}$. We separate the terms with $\cI=\{1\},\cK\backslash\{1\}$,
\be\nonumber
|\Delta^{(\cK,w)}|\Big(\sum_{a,b\in\cK}C_{ab}^2w_{ab}\Big)=\sum_{\cI\sqcup\cJ=\cK}|\Delta^{(\cI,w)}||\Delta^{(\cJ,w)}|\Big(\sum_{a\in\cI,b\in\cJ}C_{ab}w_{ab}\Big)^2
\cr
=2\sum_{\genfrac{}{}{0cm}{3}{{\genfrac{}{}{0cm}{3}{\cI\subsetneq\cK}{\cJ=\cK\backslash\cI}}}{{1\in\cI}}}\sum_{l\in\cI\backslash\{1\}}w_{1l}|\Delta^{(\cI^{\hat l}\backslash\{1\},w)}||\Delta^{(\cJ,w)}|\Big(\sum_{a\in\cI,b\in\cJ}C_{ab}w_{ab}\Big)^2+2|\Delta^{(1,w)}||\Delta^{(\cK\backslash\{1\},w)}|\Big(\sum_{b\in\cK\backslash\{1\}}C_{1b}w_{1b}\Big)^2
\ee
Note that the factor of $2$ in the second term appears to account for the possibility of $1\in\cI$ or $1\in\cJ$ and the symmetry on the sets $\cI$ and $\cJ$ of the tree splitting recursion.
Then we exchange the order of the sums first summing over $l\in\cK$ and then over the splittings of $\cK$ into $\cI$ and $\cJ$ with $l\in\cI$
\be\nonumber
|\Delta^{(\cK,w)}|=2\sum_{l\in\cK\backslash\{1\}}w_{1l}\sum_{\genfrac{}{}{0cm}{3}{{\genfrac{}{}{0cm}{3}{\cI\subsetneq\cK}{\cJ=\cK\backslash\cI}}}{{1,l\in\cI}}}|\Delta^{(\cI^{\hat l}\backslash\{1\},w)}||\Delta^{(\cJ,w)}|\frac{\Big(\sum_{a\in\cI,b\in\cJ}C_{ab}w_{ab}\Big)^2}{\sum_{a,b\in\cK}C_{ab}^2w_{ab}}+2|\Delta^{(\cK\backslash\{1\},w)}|\frac{\Big(\sum_{b\in\cK\backslash\{1\}}C_{1b}w_{1b}\Big)^2}{\sum_{a,b\in\cK}C_{ab}^2w_{ab}}
\ee
We now change the sum over splittings of $\cK$ to a sum over splittings of $\cK\backslash\{1\}$
\be\nonumber
|\Delta^{(\cK,w)}|=2\sum_{l\in\cK\backslash\{1\}}w_{1l}\sum_{\genfrac{}{}{0cm}{3}{{\genfrac{}{}{0cm}{3}{\cI\subsetneq\cK\backslash\{1\}}{\cJ=(\cK\backslash\{1\})\backslash\cI}}}{{l\in\cI}}}|\Delta^{(\cI^{\hat l},w)}||\Delta^{(\cJ,w)}|\frac{\Big(\sum_{a\in\cI^{\hat l},b\in\cJ}C_{ab}w_{ab}\Big)^2}{\sum_{a,b\in\cK}C_{ab}^2w_{ab}}
+2|\Delta^{(\cK\backslash\{1\},w)}|\frac{\Big(\sum_{b\in\cK\backslash\{1\}}C_{1b}w_{1b}\Big)^2}{\sum_{a,b\in\cK}C_{ab}^2w_{ab}}
\ee
where we have made use of the definition of the shifted weight
$$w_{i\hat l}=w_{il}+\frac{C_{i1}}{C_{il}}w_{i1}$$
to transform the sum in parenthesis in the first term into a sum over a smaller set $\cI\subsetneq\cK\backslash\{1\}$, but with the element $l\in\cI$ shifted. 

We may drop the condition $l\in\cI$ together with the overall factor of 2 using the symmetry between $\cI$ and $\cJ$. We then have a sum over splittings of $\cK^{\hat l}\backslash\{1\}$
\be\nonumber
|\Delta^{(\cK,w)}|
=\sum_{l\in\cK\backslash\{1\}}w_{1l}\sum_{\genfrac{}{}{0cm}{3}{\cI\subsetneq\cK^{\hat l}\backslash\{1\}}{\cJ=(\cK^{\hat l}\backslash\{1\})\backslash\cI}}|\Delta^{(\cI,w)}||\Delta^{(\cJ,w)}|\frac{\Big(\sum_{a\in\cI,b\in\cJ}C_{ab}w_{ab}\Big)^2}{\sum_{a,b\in\cK}C_{ab}^2w_{ab}}
+2|\Delta^{(\cK\backslash\{1\},w)}|\frac{\Big(\sum_{b\in\cK\backslash\{1\}}C_{1b}w_{1b}\Big)^2}{\sum_{a,b\in\cK}C_{ab}^2w_{ab}}
\ee

We may now use the recursion relation (\ref{rec-BG}) for $|\Delta^{(\cK^{\hat l}\backslash\{1\},w)}|$ to obtain
\be\nonumber
|\Delta^{(\cK,w)}|
=\sum_{l\in\cK\backslash\{1\}}w_{1l}|\Delta^{(\cK^{\hat l}\backslash\{1\},w)}|\frac{\sum_{a,b\in\cK^{\hat l}\backslash\{1\}}C_{ab}^2w_{ab}}{\sum_{a,b\in\cK}C_{ab}^2w_{ab}}
+2|\Delta^{(\cK\backslash\{1\},w)}|\frac{\Big(\sum_{b\in\cK\backslash\{1\}}C_{1b}w_{1b}\Big)^2}{\sum_{a,b\in\cK}C_{ab}^2w_{ab}}
\ee
Now for the recurion relation (\ref{rec-BCFW}) to hold we need
\be\nonumber
2|\Delta^{(\cK\backslash\{1\},w)}|\frac{\Big(\sum_{b\in\cK\backslash\{1\}}C_{1b}w_{1b}\Big)^2}{\sum_{a,b\in\cK}C_{ab}^2w_{ab}}=\sum_{l\in\cK\backslash\{1\}}w_{1l}|\Delta^{(\cK^{\hat l}\backslash\{1\},w)}|-\sum_{l\in\cK\backslash\{1\}}w_{1l}|\Delta^{(\cK^{\hat l}\backslash\{1\},w)}|\frac{\sum_{a,b\in\cK^{\hat l}\backslash\{1\}}C_{ab}^2w_{ab}}{\sum_{a,b\in\cK}C_{ab}^2w_{ab}}
\ee
which is equivalent to the already proven identity (\ref{ident}), which here reads
\be\nonumber
\sum_{b\in\cK\backslash\{1\}}w_{1b}C_{1b}|\Delta^{(\cK\backslash\{1\},w)}|=\sum_{l\in\cK\backslash\{1\}}w_{1l}C_{1l}|\Delta^{(\cK^{\hat l}\backslash\{1\},w)}|.
\ee
This finishes the proof.

\end{document}